\documentclass[prd,twocolumn,nopacs,floatfix,amsmath,nofootinbib,amssymb,floatfix]{revtex4}
\usepackage{graphicx,color,dcolumn,booktabs,bm}
\usepackage{longtable}
\usepackage{pdflscape}
\usepackage{pdfpages}
\usepackage{txfonts}
\usepackage{overpic}
\usepackage{amssymb}
\usepackage{makecell}
\usepackage{indentfirst}
\usepackage{feynmf}   
\usepackage{slashed}  
\usepackage{cases}
\usepackage{color}
\usepackage{multirow}
\usepackage{threeparttable}
\usepackage{epstopdf}
\usepackage{enumerate}
\usepackage{ulem}
\usepackage[figuresright]{rotating}
\usepackage[colorlinks,
            citecolor=blue,
            anchorcolor=red,
            menucolor=red,
            linkcolor=red,
            filecolor=red,
            runcolor=red,
            urlcolor=blue,
            frenchlinks=true]{hyperref}

\graphicspath{{Figures/}} %

\allowdisplaybreaks

\begin{document}

\title{Investigating topped hadrons to probe the boundaries of the potential model}
\author{Si-Qiang Luo$^{1,3,5}$}\email{luosq15@lzu.edu.cn}
\author{Qi Huang$^{2,3}$}\email{06289@njnu.edu.cn}
\author{Xiang Liu$^{1,3,4,5}$}\email{xiangliu@lzu.edu.cn}

\affiliation{
$^1$School of Physical Science and Technology, Lanzhou University, Lanzhou 730000, China\\
$^2$School of Physics and Technology, Nanjing Normal University, Nanjing 210023, China\\
$^3$Lanzhou Center for Theoretical Physics,
        Key Laboratory of Theoretical Physics of Gansu Province,
        Key Laboratory of Quantum Theory and Applications of MoE,
        Gansu Provincial Research Center for Basic Disciplines of Quantum Physics, Lanzhou University, Lanzhou 730000, China\\
        $^4$MoE Frontiers Science Center for Rare Isotopes, Lanzhou University, Lanzhou 730000, China\\
        $^5$Research Center for Hadron and CSR Physics, Lanzhou University $\&$ Institute of Modern Physics of CAS, Lanzhou 730000, China}

\begin{abstract}
Inspired by the recent discovery of a pseudoscalar enhancement structure near the $t\bar{t}$ threshold reported by the CMS and ATLAS Collaborations, this work investigates the mass spectra of single topped hadrons—including both topped mesons and topped baryons—based on a relativistic potential model. Using the same parameters obtained from the fit to mesons and baryons, we provide predictions for the mass spectra of ground and low-lying orbitally excited single topped mesons and baryons. In addition, we point out that the precise measurement of the $t\bar{t}$ mass could test the limitation of the potential model. Given the extremely large mass of the topped quark, we discuss spectroscopic properties of topped hadrons in the approximation of an ideal heavy-quark limit.
\end{abstract}
\maketitle

\section{Introduction}
Since the establishment of the quark model, theorists have attempted to explore hadronic spectra with various approaches. In the 1960s, i.e., at the beginning of the quark model, Gell-Mann and Zweig employed SU(3) flavor symmetry to classify the observed hadrons~\cite{Gell-Mann:1964ewy,Zweig:1964ruk,Zweig:1964jf}. The major success of the early quark model was that the predicted $\Omega$ baryon was confirmed by experiments~\cite{Barnes:1964pd}. In this period, the quark model was mainly based on the qualitative or semiquantitative analysis, where the targets were so-called light flavor hadrons that includeed light quarks $u$, $d$, and $s$.

In 1974, the discovery of the $J/\psi$ particle~\cite{E598:1974sol,SLAC-SP-017:1974ind} opened the door to the world of charm physics. Then, a series of charmonium was observed~\cite{Abrams:1974yy,Goldhaber:1977qn,Siegrist:1976br,Rapidis:1977cv,DASP:1978dns,Biddick:1977sv,Tanenbaum:1975ef,Whitaker:1976hb,Partridge:1980vk,Edwards:1982fif}. Based on these states, Eichten {\it et al.} established the famous Cornell model~\cite{Eichten:1974af, Eichten:1978tg, Eichten:1979ms}, where the potential contained a Coulomb-like term in addition to a linear confinement, i.e, $V(r)=-\frac{\kappa}{r} + \frac{r}{a^2}$. Since the Cornell model roughly depicted a charmonium discovered within the context of that time, it truly enabled the quantitative calculation of hadronic spectra. However, the Cornell model does not contain spin-dependent terms, which leads to a problem with this model in that it cannot be employed in spin splits. Thus, on the basis of the Cornell model, theorists developed various methods like the Isgur-Karl model~\cite{Isgur:1977ef}, Godfrey-Isgur model~\cite{Godfrey:1985xj}, and so on, where the spin splits could be well understood. In 1977, the observation of the $\Upsilon$ implied the existence of the bottom quark~\cite{E288:1977xhf}. Although its mass is much higher than that of the charm quark, the calculation implies that the potential model can also be employed in the spectra of bottom hadrons~\cite{Godfrey:1985xj}. In this way, the potential model has become a popular method of studying light falvor hadrons and heavy flavor hadrons including charm and bottom quarks, and in quiet a long time, methods based on the potential model have played a significant role in understanding hadron spectra and the inner interactions of hardons.

As the most massive of all known elementary particles, ever since the top quark was observed by the CDF and D$\O$ experiments in 1995 \cite{CDF:1995wbb,D0:1995jca}, it has attracted significant attention from the entire scientific community. The top quark not only serves as a platform for conducting precision tests of the Standard Model (SM), but also opens a window to explore new physics beyond the SM. For more details, interested readers can consult reviews~\cite{Cvetic:1997eb,Schrempp:1996fb,Atwood:2000tu,Chakraborty:2003iw,Plehn:2011tg,Quadt:2006dqn,Schilling:2012dx,Wagner:2005jh,Aguilar-Saavedra:2014kpa,Bhat:1998cd,Galtieri:2011yd,Kehoe:2007px}. Since the mean lifetime of the top quark is extremely short ($\sim 5 \times 10^{-25}$ s), it was widely accepted in the field that there is no sufficient time to bind with its antiparticle to form a bound state such as a quarkonium, as reflected in the current particle physics textbooks. This means that although the top quark was discovered many years ago, the topped hadron still remains quite mysterious. Ref.~\cite{Godfrey:1985xj} calculated the spectra of top mesons, but since the top quark was not discovered at that time, the calculation lost a crucial anchor of the top quark mass.

However, this conventional understanding has recently been greatly challenged by reports from the CMS and ATLAS Collaborations due to the observation of an enhancement structure near the $t\bar{t}$ mass threshold, which could be attributed to the possible formation of a pseudoscalar toponium \cite{CMS:2025kzt,ATLAS:2026dbe}. This new finding has naturally sparked interest among theoretical groups. For example, prior to the possible observation of the toponium, a series of studies had already predicted its existence and its properties based on perturbative quantum chromodynamics (pQCD) and potential models \cite{Fadin:1987wz,Barger:1987xg,Kuhn:1987ty,Fadin:1990wx,Strassler:1990nw,Sumino:1997ve,Hoang:2000yr,Penin:2005eu,Hagiwara:2008df,Kiyo:2008bv,Sumino:2010bv,Beneke:2015kwa,Fuks:2021xje,Garzelli:2024uhe,Wang:2024hzd,Jiang:2024fyw,Akbar:2024brg,Fuks:2024yjj,Fu:2025yft}. Their predictions align well with the experimental results. This agreement arises because, although the toponium cannot form a stable bound state due to the extremely short lifetime of the top quark, its dynamics are still governed by nonrelativistic quantum chromodynamics (NRQCD). Following the reported excess around the $t\bar{t}$ threshold, a new platform has been provided for further studies, such as spin correlations in $t\bar{t}$ production \cite{Hagiwara:2008df,Severi:2021cnj,Maltoni:2024tul,Aguilar-Saavedra:2024mnm,Nason:2025hix,Gombas:2025ibs}, developments in quantum theory \cite{Thompson:2025cgp,Lopez:2025kog,Shao:2025dzw}, and possibly explorations of new physics \cite{Butterworth:2025asm,LeYaouanc:2025mpk,Behring:2025ilo,CMS:2025dzq}. In addition, this discovery enables discussions on the potential for observing the toponium at other future colliders, such as CEPC or FCC-ee \cite{Bai:2025buy,Xiong:2025iwg}.

Over the past two decades, with the observation of a series of new hadronic states across different experiments, the study of hadron spectroscopy has entered an era of high precision \cite{Chen:2016spr,Chen:2022asf,Liu:2024uxn,Chen:2016qju,Liu:2013waa,Yuan:2018inv,Olsen:2017bmm,Guo:2017jvc,Hosaka:2016pey,Brambilla:2019esw,LHC:particles}. These findings continuously challenge the potential models. The observation of the $t\bar{t}$ enhancement structure exactly provides a good opportunity to test the potential models with extremely large quark mass. As discussed above, the potential models have achieved great success from light flavor systems to charmed states, and to bottom hadrons. If the potential models could be applied to the topped hadrons, the potential models could build a full chain to cover the hadrons including all known quarks, and the study of the topped hadron spectroscopy is the crucial part of the chain.

In this work, we employ the Godfrey-Isgur-Capstick (GIC) potential model \cite{Godfrey:1985xj,Capstick:1986ter} to study the mass spectra of singly topped hadrons, which is a relativistic quark model that has been widely used to investigate the properties of quarkonia and baryons. We point out that the precise measurement of the toponium mass is the key point to test the effectiveness of the model. In addition, because of the large mass, the topped hadrons are nearly ideal heavy-quark systems. Within the framework of heavy quark symmetry, we investigate the spectroscopic behavior of the hadronic system.

This work is organized as follows: Following this introduction, Sec. \ref{sec:model} provides a brief introduction to the GIC potential model and describes the method used to calculate the mass spectra of single topped hadrons. Our numerical results and corresponding discussion are presented in Sec. \ref{sec:results}. Then, we discuss the application of the potential in Sec. \ref{sec:applicability}. We also demonstrate the phenomenon of the system near the heavy-quark limit in Sec. \ref{sec:HQS}. Finally, in Sec. \ref{sec:summary}, the work concludes with a discussion and conclusions.

\section{Model setup}\label{sec:model}
As mentioned in the Introduction, we adopt the GIC potential model to carry on the calculation, whose Hamiltonian can be written as \cite{Godfrey:1985xj,Capstick:1986ter}
\begin{eqnarray}
H&=&\sum\limits_i\sqrt{p_i^2+m_i^2}+\sum\limits_{i<j}\left(V_{ij}^{\rm Coul}+V_{ij}^{\rm string}+V_{ij}^{\rm cont}\right.\nonumber\\
&&\left.+V_{ij}^{{\rm so}(s)}+V_{ij}^{{\rm so}(v)}+V_{ij}^{\rm tens}\right),
\end{eqnarray}
where $\sqrt{p_i^2+m_i^2}$ is the relativistic kinetic energy term, $V_{ij}^{\rm Coul}$, $V_{ij}^{\rm string}$, $V_{ij}^{\rm cont}$, $V_{ij}^{\rm so(s/v)}$, and $V_{ij}^{\rm tens}$ are the smeared semirelativistic Coulomb-type interaction, the confinement potential, the color-magnetic interaction, the spin-orbit interactions, and the tensor interaction, respectively. For the smeared semirelativistic Coulomb-type interaction, it can be explicitly expressed as
\begin{eqnarray}
    V_{ij}^{\rm Coul}=\beta_{ij}^{\frac{1}{2}+\epsilon_{\rm Coul}}\tilde{G}_{ij}\beta_{ij}^{\frac{1}{2}+\epsilon_{\rm Coul}},
\end{eqnarray}
where $\beta_{ij}=1+\frac{p_{ij}^2}{(p_{ij}^2+m_i^2)^{1/2}(p_{ij}^2+m_j^2)^{1/2}}$ is a factor that represents the relativistic effect of the interaction quarks with $p_{ij}$ being the magnitude of the momentum of either of the quarks in the $ij$ center-of-mass frame, $\epsilon_{\rm Coul}$ is a parameter used to control the magnitude of the relativistic effect in the Coulomb interaction, and $\tilde{G}_{ij}$ is the smeared Coulomb potential expressed as
\begin{eqnarray}\label{eq:smearG}
    \tilde{G}_{ij}=C_{ij}\sum\limits_k\frac{\alpha_k}{r_{ij}}{\rm erf}(\sigma_{kij} r_{ij}),
\end{eqnarray}
with $C_{ij}=-\frac{4}{3}~(-\frac{2}{3})$ being the color factor of the meson (baryon), and $\sigma_{kij}$ is defined by the smear parameter $\sigma_{ij}$ as $\sigma_{kij}^{-2}=\gamma_k^{-2}+\sigma_{ij}^{-2}$. In the definition of $\sigma_{kij}$, $\sigma_{ij}$ is parametrized by the parameters $\sigma_0$, $s$, and quark masses $m_{i/j}$ as $\sigma_{ij}^2=\sigma_0^2\left(\frac{1}{2}+\frac{1}{2}\left(\frac{4m_im_j}{(m_i+m_j)^2}\right)^4\right)+s^2\left(\frac{2m_im_j}{m_i+m_j}\right)^2$, and $\gamma_k$ is also a series of parameters, which is used to reproduce the very famous behavior of the running strong coupling constant $\alpha_s$ through $\alpha_s(Q^2)=\sum_k \alpha_k e^{-\gamma_k Q^2}$ after it is combined with another series of parameters $\alpha_k$.

Then, with the definition of $\tilde{G}_{ij}$, the color-magnetic potential $V_{ij}^{\rm cont}$, the spin-orbit interactions $V_{ij}^{{\rm so}(v)}$, and the tensor interaction $V_{ij}^{\rm tens}$ can be further represented as
\begin{eqnarray}
V_{ij}^{\rm cont}&=&\delta_{ij}^{\frac{1}{2}+\epsilon_{\rm cont}}\frac{2{\bf s}_i\cdot{\bf s}_j}{3m_im_j}\nabla^2\tilde{G}_{ij}\delta_{ij}^{\frac{1}{2}+\epsilon_{\rm cont}},\\
V_{ij}^{{\rm so}(v)}&=&\frac{1}{r_{ij}}\frac{{\rm d}\tilde{G}_{ij}}{{\rm d}r_{ij}}\left(
 \delta_{ii}^{\frac{1}{2}+\epsilon_{{\rm so}(v)}}\frac{{\bf r}_{ij}\times{\bf p}_i\cdot{\bf s}_i}{2m_i^2}\delta_{ii}^{\frac{1}{2}+\epsilon_{{\rm so}(v)}}\right.\nonumber\\
&&\left.-\delta_{jj}^{\frac{1}{2}+\epsilon_{{\rm so}(v)}}\frac{{\bf r}_{ij}\times{\bf p}_j\cdot{\bf s}_j}{2m_j^2}\delta_{jj}^{\frac{1}{2}+\epsilon_{{\rm so}(v)}}\right.\nonumber\\
&&\left.-\delta_{ij}^{\frac{1}{2}+\epsilon_{{\rm so}(v)}}\frac{{\bf r}_{ij}\times{\bf p}_j\cdot{\bf s}_i-{\bf r}_{ij}\times{\bf p}_i\cdot{\bf s}_j}{2m_i^2}\delta_{ij}^{\frac{1}{2}+\epsilon_{{\rm so}(v)}}\right),\\
V_{ij}^{\rm tens}&=&\delta_{ij}^{\frac{1}{2}+\epsilon_{\rm tens}}\frac{1}{3m_im_j}\left(\frac{3({\bf s}_i\cdot{\bf r}_{ij})({\bf s}_j\cdot{\bf r}_{ij})}{r_{ij}^2}-{\bf s}_i\cdot{\bf s}_j\right)\nonumber\\
&&\times\left(\frac{1}{r_{ij}}\frac{{\rm d}\tilde{G}_{ij}}{{\rm d}r_{ij}}-\frac{{\rm d}^2\tilde{G}_{ij}}{{\rm d}r_{ij}^2}\right)\delta_{ij}^{\frac{1}{2}+\epsilon_{\rm tens}},
\end{eqnarray}
where $\delta_{ij}=\frac{m_im_j}{(p_{ij}^2+m_i^2)^{1/2}(p_{ij}^2+m_j^2)^{1/2}}$ and $\epsilon_{\rm kind}$ are still used to control the magnitude of the relativistic effect in potential $V_{ij}^{\rm kind}$.

Finally, for the flavor-independent linear-type confinement, its smeared form is
\begin{eqnarray}\label{eq:confinement}
V_{ij}^{\rm string}&=&\int{\rm d}^3 {\bf r}^\prime \left(\frac{\sigma_{ij}^3}{\pi^{3/2}}{\rm e}^{-\sigma_{ij}^2({\bf r}-{\bf r}^\prime)^2}\right)\left(-\frac{3}{4}C_{ij}(br_{ij}+c)\right)\nonumber\\
&=&-\frac{3}{4}C_{ij}c-\frac{3}{4}C_{ij}br_{ij}\left(\frac{{\rm e}^{-\sigma_{ij}^2r_{ij}^2}}{\sqrt{\pi}\sigma_{ij}r_{ij}}\vphantom{+\left(1+\frac{1}{2\sigma_{ij}^2r_{ij}^2}\right){\rm erf}(\sigma_{ij}r_{ij})}\right.\nonumber\\
&&\left.\vphantom{\frac{{\rm e}^{-\sigma_{ij}^2r_{ij}^2}}{\sqrt{\pi}\sigma_{ij}r_{ij}}}+\left(1+\frac{1}{2\sigma_{ij}^2r_{ij}^2}\right){\rm erf}(\sigma_{ij}r_{ij})\right),
\end{eqnarray}
with which the Thomas precession term $V_{ij}^{{\rm so}(s)}$ can then be expressed by the confinement as
\begin{eqnarray}
V_{ij}^{{\rm so}(s)}&=&\frac{1}{r_{ij}}\frac{{\rm d}V_{ij}^{\rm string}}{{\rm d}r_{ij}}\left(
-\delta_{ii}^{\frac{1}{2}+\epsilon_{{\rm so}(s)}}\frac{{\bf r}_{ij}\times{\bf p}_i\cdot{\bf s}_i}{2m_i^2}\delta_{ii}^{\frac{1}{2}+\epsilon_{{\rm so}(s)}}\right.\nonumber\\
&&\left.+\delta_{jj}^{\frac{1}{2}+\epsilon_{{\rm so}(s)}}\frac{{\bf r}_{ij}\times{\bf p}_j\cdot{\bf s}_j}{2m_j^2}\delta_{jj}^{\frac{1}{2}+\epsilon_{{\rm so}(s)}}\right).
\end{eqnarray}

With the above potentials, we can solve the stationary state Schr\"odinger equation to get the single topped hadron spectra, which is symbolically given as
\begin{eqnarray}
    H|\psi\rangle = E|\psi\rangle.
\end{eqnarray}
Here, the wave function $|\psi\rangle$ can be written as a direct product of color, spin-$S$, flavor, and orbit-$L$ wave functions as
\begin{eqnarray}
    \psi=\mathcal{A}\left[\psi^{\rm color} \otimes \psi^{\rm flavor} \otimes \left[\psi_S^{\rm spin} \otimes \psi_L^{\rm orbit}\right]_J\right],  
\end{eqnarray}
with $\mathcal{A}$ being the operator to make the whole wave function $\psi$ antisymmetric, and $J$ is the total angular momentum of the system. 

For the color wave function $\psi^{\rm color}$, when it is a meson or a baryon, its explicit form is
\begin{eqnarray}
    \psi^{\rm color}_{\rm meson} &=& \frac{1}{\sqrt{3}}\left(r\bar{r}+g\bar{g}+b\bar{b}\right),\\
    \psi^{\rm color}_{\rm baryon} &=&\frac{1}{\sqrt{6}}\left(r g b-g r b+b r g-r b g+g b r-b g r\right).
\end{eqnarray}
For the flavor wave functions $\psi^{\rm flavor}$ of specific single topped hadrons, their expressions are as follows
\begin{eqnarray}
    &&T_n=t\bar{n}~(n=u,~d),\quad T_s=t\bar{s},\quad T_c=t\bar{c},\quad T_b=t\bar{b},\nonumber\\
    &&\Lambda_t = \frac{1}{\sqrt{2}}\left(ud-du\right)t,\quad \Xi_t = \frac{1}{\sqrt{2}}\left(ns-sn\right)t,\nonumber\\
    &&\Xi_{nct} = \frac{1}{\sqrt{2}}\left(nc-cn\right)t,\quad\Xi_{sct} = \frac{1}{\sqrt{2}}\left(sc-cs\right)t,\nonumber\\
    &&\Xi_{nbt} = \frac{1}{\sqrt{2}}\left(nb-bn\right)t,\quad\Xi_{sbt} = \frac{1}{\sqrt{2}}\left(sb-bs\right)t,\nonumber\\
    &&\Xi_{cbt} = \frac{1}{\sqrt{2}}\left(cb-bc\right)t,\quad\Sigma_t = \frac{1}{\sqrt{2}}\left(ud+du\right)t,\nonumber\\
    &&\Xi_t' = \frac{1}{\sqrt{2}}\left(ns+sn\right)t,\quad\Xi_{nct}' = \frac{1}{\sqrt{2}}\left(nc+cn\right)t,\nonumber\\
    &&\Xi_{sct}' = \frac{1}{\sqrt{2}}\left(sc+cs\right)t,\quad\Xi_{nbt}' = \frac{1}{\sqrt{2}}\left(nb+bn\right)t,\nonumber\\
    &&\Xi_{sbt}' = \frac{1}{\sqrt{2}}\left(sb+bs\right)t,\quad\Xi_{cbt}' = \frac{1}{\sqrt{2}}\left(cb+bc\right)t,\nonumber\\
    &&\Omega_{t} = sst,\quad\Omega_{cct} = cct,\quad\Omega_{bbt} = bbt.
\end{eqnarray}
Then, for the spin-$S$ and orbit-$L$ wave functions $\psi_S^{\rm spin}$ and $\psi_L^{\rm orbit}$, their expressions can be generally constructed by the Clebsch-Gordan series as
\begin{eqnarray}
    &&\psi_S^{\rm spin} = \left[\left[\left[\psi_{S_1}^{\rm spin}\otimes\psi_{S_2}^{\rm spin}\right]_{S_{1,2}}\otimes \psi_{S_3}^{\rm spin}\right]_{S_{12,3}}\otimes...\right]_S,\\
    &&\psi_L^{\rm orbit} = \left[\left[\left[\psi_{L_{1,2}}^{\rm orbit}\otimes\psi_{L_{12,3}}^{\rm orbit}\right]_{L_{123}}\otimes \psi_{L_{123,4}}^{\rm orbit}\right]_{L_{1234}}\otimes...\right]_L,
\end{eqnarray}
where $S_i$ means the spin of cluster $i$, $S_{i,j}$ is the total spin coupled by clusters $i$ and $j$, $L_{i,j}$ denotes the relative orbital angular momentum between clusters $i$ and $j$, and $L_{ij}$ represents the total orbital angular momentum coupled by clusters $i$ and $j$. To get the spectra through the Rayleigh-Ritz variation method, for each relative orbit angular momentum wave function, we expand it into a series of basis as $\psi^{\rm orbit}_l = \sum\limits_{n=1}^{N_{max}} C_n \psi^{\rm orbit}_{nl}$. As for the expansion basis, in this work, we adopt the Gaussian function, which is proved to be convenient and effective in handling the few-body problem \cite{Hiyama:2003cu} as
\begin{eqnarray}
    \psi^{\rm orbit}_{nl} = \begin{cases}N_{nl}^r e^{-\nu_n r^2}\mathcal{Y}_l(\boldsymbol{r}),~\mathrm{for~coordinate~representation,}\\
    N_{nl}^p e^{-\frac{p^2}{4\nu_n}}\mathcal{Y}_l(\boldsymbol{p}),~\mathrm{for~momentum~representation,}\end{cases}
\end{eqnarray}
with $\nu_n$ being parametrized by two parameters $r_{min}$ and $r_{max}$ as $\nu_n=r_{min}^{-2}(r_{max}/r_{min})^{(2-2n)/n_{max}}$, and $N_{nl}^{r/p}$ is the normalization factor written as
\begin{eqnarray}
    N_{nl}^r &=& \left(\frac{2^{l+2}\left(2 v_n\right)^{l+\frac{3}{2}}}{\sqrt{\pi}(2 l+1)!!}\right)^{\frac{1}{2}},\\
    N_{nl}^p &=& (-i)^l\left(\frac{2^{l+2}\left(2 v_n\right)^{-l-\frac{3}{2}}}{\sqrt{\pi}(2 l+1)!!}\right)^{\frac{1}{2}}.
\end{eqnarray}

Finally, to exhibit the heavy-quark symmetry, we transform the wave function $\psi$ from $S$-$L$ coupling to the $j_\ell$-$J$ representation, where $J$ is the total angular momentum, and $j_\ell=J\pm S_t$ denotes the quantum number of the light degree of freedom ($S_t=\frac{1}{2}$ is the spin of the top quark) as
\begin{eqnarray}
    \left[\psi_{j_\ell}\otimes\psi_{S_t}^{\rm spin}\right]_J &=& (-1)^{L+S_{t\!\!/}+J+\frac{1}{2}} \sum\limits_S\sqrt{2 j_\ell+1}\nonumber\\
    &&\times\sqrt{2 S+1}\left\{\begin{array}{ccc}
    L & S_{t\!\!/} & j_\ell \\
    S_t & J & S
    \end{array}\right\}\nonumber\\
    &&\times\left[\psi_S^{\rm spin} \otimes \psi_L^{\rm orbit}\right]_J,
\end{eqnarray}
where $S_{t\!\!/}$, in single topped hadrons, means the total spin composed by other quarks except the top quark.

\section{Numerical results}\label{sec:results}

\begin{table}[htbp]
\caption{The parameters used in this work, in which the superscript “$\ast$” denotes the parameters that were fixed during the fitting process. For the meson sector parameters, we adopt the values from Ref.~\cite{Godfrey:1985xj}.}
\label{tab:widthsN2}
\renewcommand\arraystretch{1.3}
\centering
\begin{tabular*}{86mm}{@{\extracolsep{\fill}}lcc}
\toprule[1.00pt]
\toprule[1.00pt]
Parameters               &Meson~\cite{Godfrey:1985xj}                              &Baryon                             \\
\midrule[0.75pt]
$m_n$ (GeV)              &\multicolumn{2}{c}{$0.220^\ast$}                                                             \\
$m_s$ (GeV)              &\multicolumn{2}{c}{$0.419^\ast$}                                                             \\
$m_c$ (GeV)              &\multicolumn{2}{c}{$1.628^\ast$}                                                             \\
$m_b$ (GeV)              &\multicolumn{2}{c}{$4.977^\ast$}                                                             \\
$m_t$ (GeV)              &\multicolumn{2}{c}{$172.57$~\cite{ParticleDataGroup:2024cfk}}                                \\
$\alpha_k$               &\multicolumn{2}{c}{$\left[0.25^\ast,~0.15^\ast,~0.20^\ast\right]$}                           \\
$\gamma_k$               &\multicolumn{2}{c}{$\left[\frac{1}{2}^\ast, \sqrt{\frac{5}{2}}^\ast,5\sqrt{10}^\ast\right]$} \\
$b$ (${\rm GeV}^2$)      & 0.18$^\ast$                                             & 0.141                             \\
$c$ (GeV)                &-0.253$^\ast$                                            &-0.204                             \\
$\sigma_0$ (GeV)         & 1.80$^\ast$                                             & 1.889                             \\
$s$                      & 1.55$^\ast$                                             & 1.422                             \\
$\epsilon_{\rm cont}$    &-0.168$^\ast$                                            &-0.156                             \\
$\epsilon_{\rm tens}$    &+0.025$^\ast$                                            &-0.379                             \\
$\epsilon_{{\rm so}(v)}$ &-0.035$^\ast$                                            &+0.006                             \\
$\epsilon_{{\rm so}(s)}$ &+0.055$^\ast$                                            &+0.449                             \\
$\epsilon_{\rm Coul}$    & 0$^\ast$                                                & 0$^\ast$                          \\
\bottomrule[1.00pt]
\bottomrule[1.00pt]
\end{tabular*}
\end{table}

\begin{table*}[htbp]
\caption{The comparisons of the theoretical and experimental masses of the charm and bottom baryons. We also present the $\chi^2/{ d.o.f.}$ value for parameter fitting. Here, the results of theoretical mass $M^{\rm The.}$, experimental result $M^{\rm Exp.}$ obtained from Ref.~\cite{ParticleDataGroup:2024cfk}, and experimental error $M^{\rm Err.}$ in Ref.~\cite{ParticleDataGroup:2024cfk} are in units of MeV.}
\label{tab:fbaryon}
\renewcommand\arraystretch{1.3}
\centering
\begin{tabular*}{180mm}{@{\extracolsep{\fill}}lcllcll}
\toprule[1.00pt]
\toprule[1.00pt]
\multirow{2}{*}{States} &\multicolumn{3}{c}{Charm }                                                                                                                               &\multicolumn{3}{c}{Bottom}                                                                                                                               \\
\Xcline{2-4}{0.75pt}
\Xcline{5-7}{0.75pt}
                        &$M^{\rm The.}$ &\multicolumn{1}{c}{$M^{\rm Exp.}$~\cite{ParticleDataGroup:2024cfk}} &\multicolumn{1}{c}{$M^{\rm Err.}$~\cite{ParticleDataGroup:2024cfk}} &$M^{\rm The.}$ &\multicolumn{1}{c}{$M^{\rm Exp.}$~\cite{ParticleDataGroup:2024cfk}} &\multicolumn{1}{c}{$M^{\rm Err.}$~\cite{ParticleDataGroup:2024cfk}} \\
\midrule[0.75pt]
$\Lambda_Q(1S)$         &2287.44        &2286.46                                                             &0.14                                                                & 5621.63       &5619.60                                                             &0.17                                                                \\
$\Lambda_Q(2S)$         &2763.16        &2766.6                                                              &2.4                                                                 & 6042.17       &6072.3                                                              &2.9                                                                 \\
$\Lambda_Q(1P,1/2^-)$   &2606.47        &2592.25                                                             &0.28                                                                & 5903.02       &5912.19                                                             &0.17                                                                \\
$\Lambda_Q(1P,3/2^-)$   &2628.01        &2628.00                                                             &0.15                                                                & 5911.97       &5912.01                                                             &0.17                                                                \\
$\Lambda_Q(1D,3/2^+)$   &2878.78        &2856.1                                                              &4.15                                                                & 6138.52       &6146.2                                                              &0.4                                                                 \\
$\Lambda_Q(1D,5/2^+)$   &2891.72        &2881.63                                                             &0.24                                                                & 6145.67       &6152.5                                                              &0.4                                                                 \\
$\Sigma_Q(1S)$          &2447.14        &2453.97                                                             &0.14                                                                & 5811.60       &5810.56                                                             &0.25                                                                \\
$\Sigma_Q^*(1S)$        &2524.30        &2518.41                                                             &0.22                                                                & 5839.81       &5830.32                                                             &0.27                                                                \\
$\Xi_Q(1S)$             &2475.63        &2467.71                                                             &0.23                                                                & 5803.15       &5791.9                                                              &0.5                                                                 \\
$\Xi_Q(2S)$             &2945.59        &2964.3                                                              &1.5                                                                 & 6221.57       &$\cdots$                                                            &$\cdots$                                                            \\
$\Xi_Q(1P,1/2^-)$       &2793.46        &2791.9                                                              &0.5                                                                 & 6084.41       &6087.2                                                              &0.5                                                                 \\
$\Xi_Q(1P,3/2^-)$       &2813.84        &2816.51                                                             &0.25                                                                & 6093.05       &6099.8                                                              &0.6                                                                 \\
$\Xi_Q(1D,3/2^+)$       &3062.59        &3055.9                                                              &0.4                                                                 & 6317.44       &6327.28                                                             &0.35                                                                \\
$\Xi_Q(1D,5/2^+)$       &3073.83        &3077.2                                                              &0.4                                                                 & 6323.99       &6332.69                                                             &0.28                                                                \\
$\Xi_Q^\prime(1S)$      &2581.54        &2578.2                                                              &0.5                                                                 & 5937.06       &5935.1                                                              &0.5                                                                 \\
$\Xi_Q^*(1S)$           &2651.83        &2645.10                                                             &0.30                                                                & 5963.15       &5955.7                                                              &0.5                                                                 \\
$\Omega_Q(1S)$          &2689.86        &2695.2                                                              &1.7                                                                 & 6037.07       &6045.8                                                              &0.8                                                                 \\
$\Omega_Q^*(1S)$        &2757.73        &2765.9                                                              &2.0                                                                 & 6062.63       &$\cdots$                                                            &$\cdots$                                                            \\
$\Xi_{QQ}(1S)$          &3612.30        &3621.6                                                              &0.4                                                                 &10174.92       &$\cdots$                                                            &$\cdots$                                                            \\
\midrule[0.75pt]
\multicolumn{7}{c}{$\chi^2/d.o.f.\approx678$}                                                                                                                                                                                                                                                                                               \\
\bottomrule[1.00pt]
\bottomrule[1.00pt]
\end{tabular*}
\end{table*}

To demonstrate the consistency of the GIC potential model \cite{Godfrey:1985xj,Capstick:1986ter}, we begin by determining the model parameters. For mesons, we directly adopt the parameters from Ref.~\cite{Godfrey:1985xj}, while for baryons, we perform a fit using the well-established charmed and bottomed baryons. In our numerical calculations of baryon systems, we use $n_{\text{max}} = 10$ Gaussian basis functions with $r_{\text{min}} = 0.1$ fm and $r_{\text{max}} = 3.0$ fm. The fitted model parameters are presented in Table~\ref{tab:widthsN2}, and the corresponding baryon spectra obtained in the fit are compiled in Table~\ref{tab:fbaryon}. The results indicate that the model provides a satisfactory description of the available experimental data on low-lying heavy baryons. We therefore adopt the parameter set from Table~\ref{tab:widthsN2} for all subsequent calculations.

As an initial test, we compute the $S$-wave toponium spectrum. The constituent mass of the top quark is set to $m_t = 172.57$ GeV, consistent with the value reported by the Review of Particle Physics \cite{ParticleDataGroup:2024cfk}. This choice is motivated by the argument in Ref.~\cite{Binosi:2022djx} that the constituent mass approaches the current mass as the quark becomes heavier. Given the large current mass of the top quark, the strong coupling constant $\alpha_s$ is very small in this regime, and the Schwinger mechanism is expected to play a negligible role. Therefore, equating the constituent and current masses of the top quark in our potential model calculation is well justified.

As a result, the masses of the ground pseudoscalar and vector toponium are
\begin{eqnarray}
    &&m_{t\bar{t}}\left(^1S_0\right) = 342.618~\mathrm{GeV},\\
    &&m_{t\bar{t}}\left(^3S_1\right) = 342.628~\mathrm{GeV}.
\end{eqnarray}
Our result for the pseudoscalar toponium is in excellent agreement with the location of the enhancement structure set in the CMS and ATLAS experiments \cite{CMS:2025kzt,ATLAS:2026dbe}. Moreover, due to the extremely large mass of the top quark, the masses of the ground $^1S_0$ and $^3S_1$ states are nearly degenerate. This can be attributed to the color-magnetic interaction, whose contribution is inversely proportional to the quark masses. The consistency between our theoretical result and the experimental observation strongly supports the direct extension of the GIC potential model and our fitted parameters to the top quark region.

\begin{figure*}[htbp]
    \centering
    \includegraphics[width=\textwidth]{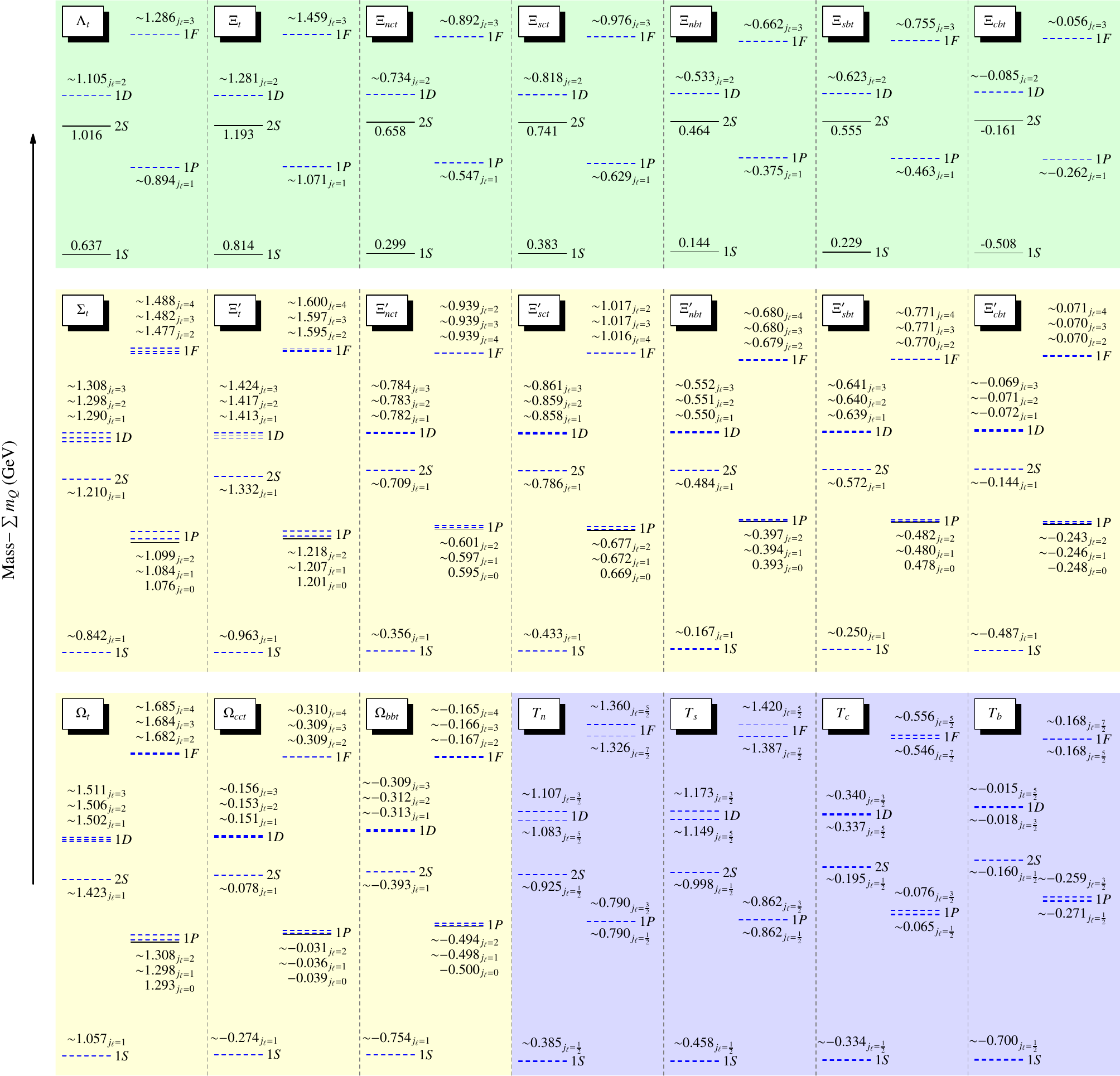}
    \caption{The mass spectra of single topped baryons and mesons are presented, with green and yellow panels indicating baryon spectra, and purple panels representing meson spectra. Solid lines denote the presence of a single state, while dashed lines indicate two or more quasidegenerate states in the corresponding region. To more clearly display the mass splittings, the heavy flavor quark masses ($\sum m_Q,~Q=c,b,t$) have been subtracted.}
    \label{fig:spectrum}
\end{figure*}

Encouraged by this agreement, we proceed to provide predictions for the mass spectra of low-lying radially and orbitally excited single topped baryons and mesons. The complete spectra up to the $1F$ states are presented in Fig.~\ref{fig:spectrum}, while the detailed numerical values for single topped mesons and baryons are compiled in Tables~\ref{tab:tmeson} and \ref{tab:tbaryon}, respectively. 

\begin{table}[htbp]
\caption{The subtracted mass spectra of the single topped mesons that correspond to Fig.~\ref{fig:spectrum} in units of GeV.}
\label{tab:tmeson}
\renewcommand\arraystretch{1.15}
\centering
\begin{tabular*}{86mm}{@{\extracolsep{\fill}}lcccc}
\toprule[1.00pt]
\toprule[1.00pt]
States                                &$T_n$  &$T_s$  &$T_c$  &$T_b$  \\
\midrule[0.75pt]
$|1S,0^-\rangle$                      & 0.384 & 0.457 &-0.335 &-0.702 \\
$|2S,0^-\rangle$                      & 0.924 & 0.997 & 0.194 &-0.161 \\
$|1S,1^-\rangle$                      & 0.386 & 0.459 &-0.332 &-0.698 \\
$|2S,1^-\rangle$                      & 0.925 & 0.998 & 0.195 &-0.159 \\
$|1P,0^+\rangle_{j_\ell=\frac{1}{2}}$ & 0.789 & 0.861 & 0.064 &-0.271 \\
$|1P,1^+\rangle_{j_\ell=\frac{1}{2}}$ & 0.790 & 0.862 & 0.065 &-0.270 \\
$|1P,1^+\rangle_{j_\ell=\frac{3}{2}}$ & 0.789 & 0.862 & 0.076 &-0.260 \\
$|1P,2^+\rangle_{j_\ell=\frac{3}{2}}$ & 0.790 & 0.862 & 0.077 &-0.259 \\
$|1D,1^-\rangle_{j_\ell=\frac{3}{2}}$ & 1.107 & 1.173 & 0.340 &-0.019 \\
$|1D,2^-\rangle_{j_\ell=\frac{3}{2}}$ & 1.108 & 1.173 & 0.341 &-0.018 \\
$|1D,2^-\rangle_{j_\ell=\frac{5}{2}}$ & 1.082 & 1.149 & 0.337 &-0.016 \\
$|1D,3^-\rangle_{j_\ell=\frac{5}{2}}$ & 1.083 & 1.149 & 0.338 &-0.015 \\
$|1F,2^+\rangle_{j_\ell=\frac{5}{2}}$ & 1.360 & 1.420 & 0.556 & 0.168 \\
$|1F,3^+\rangle_{j_\ell=\frac{5}{2}}$ & 1.360 & 1.420 & 0.556 & 0.168 \\
$|1F,3^+\rangle_{j_\ell=\frac{7}{2}}$ & 1.326 & 1.387 & 0.546 & 0.168 \\
$|1F,4^+\rangle_{j_\ell=\frac{7}{2}}$ & 1.326 & 1.387 & 0.546 & 0.168 \\
\bottomrule[1.00pt]
\bottomrule[1.00pt]
\end{tabular*}
\end{table}

\begin{table*}[htbp]
\caption{The subtracted mass spectra of the single topped baryons that correspond to Fig.~\ref{fig:spectrum} in units of GeV.}
\label{tab:tbaryon}
\renewcommand\arraystretch{1.3}
\centering
\begin{tabular*}{180mm}{@{\extracolsep{\fill}}lcccccccccc}
\toprule[1.00pt]
\toprule[1.00pt]
States                          &$\Lambda_t$ &$\Xi_t$        &$\Xi_{nct}$        &$\Xi_{sct}$        &$\Xi_{nbt}$        &$\Xi_{sbt}$        &$\Xi_{cbt}$        &           &               &               \\
\midrule[0.75pt]
$|1S,1/2^+\rangle$              & 0.637      & 0.814         & 0.299             & 0.383             & 0.144             & 0.229             &-0.508             &           &               &               \\
$|2S,1/2^+\rangle$              & 1.016      & 1.193         & 0.658             & 0.741             & 0.464             & 0.555             &-0.161             &           &               &               \\
$|1P,1/2^-\rangle$              & 0.893      & 1.071         & 0.547             & 0.629             & 0.375             & 0.463             &-0.262             &           &               &               \\
$|1P,3/2^-\rangle$              & 0.894      & 1.071         & 0.547             & 0.629             & 0.375             & 0.463             &-0.261             &           &               &               \\
$|1D,3/2^+\rangle$              & 1.105      & 1.281         & 0.734             & 0.817             & 0.533             & 0.623             &-0.085             &           &               &               \\
$|1D,5/2^+\rangle$              & 1.105      & 1.281         & 0.734             & 0.818             & 0.533             & 0.623             &-0.085             &           &               &               \\
$|1F,5/2^-\rangle$              & 1.286      & 1.459         & 0.892             & 0.976             & 0.662             & 0.755             & 0.056             &           &               &               \\
$|1F,7/2^-\rangle$              & 1.286      & 1.459         & 0.892             & 0.976             & 0.662             & 0.755             & 0.056             &           &               &               \\
\bottomrule[0.75pt]
\toprule[0.75pt]
States                          &$\Sigma_t$  &$\Xi_t^\prime$ &$\Xi_{nct}^\prime$ &$\Xi_{sct}^\prime$ &$\Xi_{nbt}^\prime$ &$\Xi_{sbt}^\prime$ &$\Xi_{cbt}^\prime$ &$\Omega_t$ &$\Omega_{cct}$ &$\Omega_{bbt}$ \\
\midrule[0.75pt]
$|1S,1/2^+\rangle$              & 0.841      & 0.962         & 0.355             & 0.433             & 0.166             & 0.249             &-0.487             & 1.057     &-0.275         &-0.754         \\
$|2S,1/2^+\rangle$              & 1.210      & 1.332         & 0.709             & 0.786             & 0.483             & 0.572             &-0.144             & 1.423     & 0.078         &-0.393         \\
$|1S,3/2^+\rangle$              & 0.842      & 0.963         & 0.356             & 0.434             & 0.167             & 0.250             &-0.486             & 1.058     &-0.274         &-0.753         \\
$|2S,3/2^+\rangle$              & 1.211      & 1.332         & 0.710             & 0.786             & 0.484             & 0.572             &-0.143             & 1.423     & 0.078         &-0.393         \\
$|1P,1/2^-\rangle_{j_{\ell}=0}$ & 1.076      & 1.201         & 0.595             & 0.669             & 0.393             & 0.478             &-0.248             & 1.293     &-0.039         &-0.500         \\
$|1P,1/2^-\rangle_{j_{\ell}=1}$ & 1.083      & 1.207         & 0.597             & 0.672             & 0.394             & 0.479             &-0.247             & 1.298     &-0.037         &-0.498         \\
$|1P,3/2^-\rangle_{j_{\ell}=1}$ & 1.084      & 1.207         & 0.597             & 0.672             & 0.395             & 0.480             &-0.246             & 1.298     &-0.036         &-0.498         \\
$|1P,3/2^-\rangle_{j_{\ell}=2}$ & 1.098      & 1.218         & 0.601             & 0.677             & 0.397             & 0.482             &-0.243             & 1.307     &-0.031         &-0.495         \\
$|1P,5/2^-\rangle_{j_{\ell}=2}$ & 1.099      & 1.218         & 0.602             & 0.677             & 0.397             & 0.483             &-0.243             & 1.308     &-0.031         &-0.494         \\
$|1D,1/2^+\rangle_{j_{\ell}=1}$ & 1.290      & 1.412         & 0.782             & 0.858             & 0.550             & 0.639             &-0.072             & 1.502     & 0.151         &-0.313         \\
$|1D,3/2^+\rangle_{j_{\ell}=1}$ & 1.290      & 1.413         & 0.782             & 0.858             & 0.550             & 0.639             &-0.072             & 1.502     & 0.151         &-0.313         \\
$|1D,3/2^+\rangle_{j_{\ell}=2}$ & 1.298      & 1.417         & 0.783             & 0.859             & 0.551             & 0.640             &-0.071             & 1.506     & 0.153         &-0.312         \\
$|1D,5/2^+\rangle_{j_{\ell}=2}$ & 1.298      & 1.417         & 0.783             & 0.860             & 0.551             & 0.640             &-0.071             & 1.506     & 0.153         &-0.311         \\
$|1D,5/2^+\rangle_{j_{\ell}=3}$ & 1.308      & 1.423         & 0.784             & 0.861             & 0.552             & 0.641             &-0.069             & 1.511     & 0.155         &-0.310         \\
$|1D,7/2^+\rangle_{j_{\ell}=3}$ & 1.308      & 1.424         & 0.784             & 0.861             & 0.553             & 0.641             &-0.069             & 1.511     & 0.156         &-0.309         \\
$|1F,3/2^-\rangle_{j_{\ell}=2}$ & 1.476      & 1.594         & 0.939             & 1.017             & 0.679             & 0.770             & 0.069             & 1.682     & 0.309         &-0.167         \\
$|1F,5/2^-\rangle_{j_{\ell}=2}$ & 1.477      & 1.595         & 0.939             & 1.017             & 0.679             & 0.770             & 0.070             & 1.682     & 0.309         &-0.167         \\
$|1F,5/2^-\rangle_{j_{\ell}=3}$ & 1.482      & 1.597         & 0.939             & 1.017             & 0.680             & 0.771             & 0.070             & 1.683     & 0.309         &-0.166         \\
$|1F,7/2^-\rangle_{j_{\ell}=3}$ & 1.482      & 1.597         & 0.939             & 1.017             & 0.680             & 0.771             & 0.070             & 1.684     & 0.310         &-0.166         \\
$|1F,7/2^-\rangle_{j_{\ell}=4}$ & 1.488      & 1.600         & 0.939             & 1.016             & 0.680             & 0.771             & 0.071             & 1.685     & 0.310         &-0.165         \\
$|1F,9/2^-\rangle_{j_{\ell}=4}$ & 1.489      & 1.600         & 0.939             & 1.016             & 0.680             & 0.771             & 0.071             & 1.685     & 0.310         &-0.165         \\
\bottomrule[1.00pt]
\bottomrule[1.00pt]
\end{tabular*}
\end{table*}

Given the large mass of the top quark, we subtract its mass in Fig.~\ref{fig:spectrum}. In addition, to present the mass gap more clearly and highlight the underlying physical features, we also subtract the masses of the $b$ and $c$ quarks. As for light flavor quarks, since they contain large relativistic effects in such systems and consequently affect the mass spectra a lot, their masses are retained in our numerical results. Then, a striking observation emerges: Although the specific spectral values differ among various systems, the mass splittings between states of different quantum numbers exhibit a remarkably consistent behavior, underscoring the robustness of heavy-quark symmetry.

Moreover, from the ground state masses of single topped mesons listed in Table~\ref{tab:tmeson}, we observe that the binding energy follows the hierarchy $\Delta E_{T_b} < \Delta E_{T_c} < \Delta E_{T_s} < \Delta E_{T_n}$. Given that the running coupling constant $\alpha_s$ follows the same ordering across these systems, this trend suggests that lighter quarks must experience stronger relativistic effects—such as effective mass increase—to form quasibound states with the top quark.

Furthermore, both Fig.~\ref{fig:spectrum} and Tables~\ref{tab:tmeson} and~\ref{tab:tbaryon} reveal that within any specific system, states sharing the same light degrees of freedom exhibit near-degenerate masses. This can be understood from the fact that all noncentral potentials are inversely proportional to the quark masses; hence, their contributions are strongly suppressed by the heavy top quark. A similar phenomenon has also been observed in QCD sum rule calculations \cite{Zhang:2025xxd,Zhang:2025fdp}.

\section{The applicability of the potential models}\label{sec:applicability}

\begin{figure}
    \centering
    \includegraphics[width=86mm]{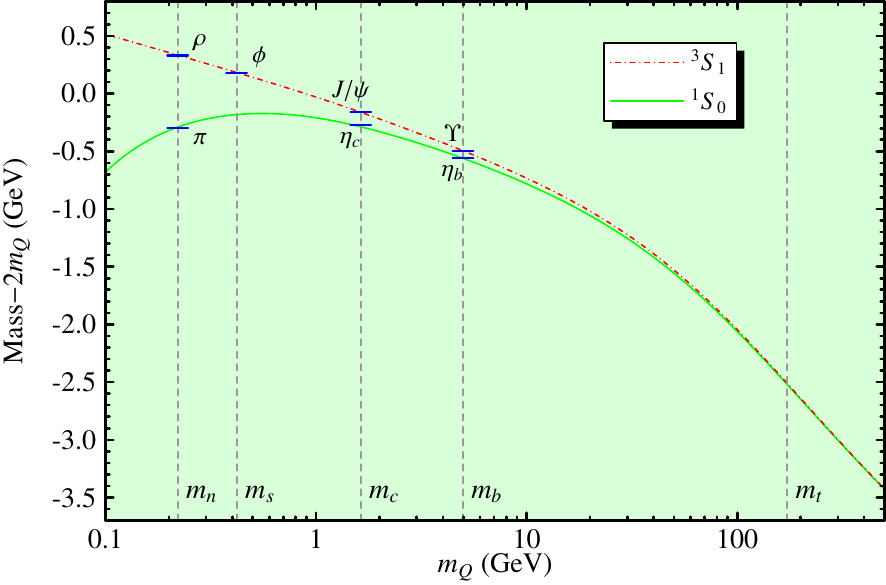}
    \caption{The mass spectra of $^3S_1$ and $^1S_0$ quarkonia as functions of the quark mass $m_Q$. The short blue lines are observed $^3S_1$ and $^1S_0$ states.}
    \label{fig:spectrum_QQ}
\end{figure}

In Ref.~\cite{Godfrey:1985xj}, with the observed abundant mesons, the effectiveness of the GIC model has been well employed in meson systems, from the lightest $\pi$ to the bottom quarkonia. In Ref.~\cite{Capstick:1986ter}, the model was employed in baryonic systems, from light flavor to singly heavy flavor baryons, but during the period of Ref.~\cite{Capstick:1986ter}, most singly heavy flavor baryons were not observed. As the experiments progressed, a series of excited state observations suggested that the GIC model could also be applied to singly heavy baryon systems. However, when we extend the potential model to the top quark, there is a crucial problem, i.e., we do not know if the potential model could be employed in the topped hadrons. In Fig.~\ref{fig:spectrum_QQ}, we present the spectra of $^3S_1$ and $^1S_0$ quarkonia with quark mass $m_Q$ in the range of 0.1$\sim$500 GeV, which covers light flavor mesons to topped quarkonia. For the observed states, the calculated masses could well match the measured results for both $^3S_1$ and $^1S_0$ states. Here, there exists a large mass gap between the bottom and top quarks. Although CMS and ATLAS observed the enhancement structure near the $t\bar{t}$ threshold \cite{CMS:2025kzt,ATLAS:2026dbe}, precisely measuring the mass of $t\bar{t}$ is still challenging work. According to Fig.~\ref{fig:spectrum_QQ},  the masses of $t\bar{t}$ with subtraction of $2m_t$ are about $-2.5$ GeV for both $^3S_1$ and $^1S_0$. We notice that this value is much higher than the states composed of $u$, $d$, $s$, $c$, and $b$. Further study implies that this value highly depends on the smeared Coulomb potential in Eq.~(\ref{eq:smearG}) and constant $c$ of the confinement in Eq.~(\ref{eq:confinement}). If $\alpha_s$ is taken as a constant, the solution of the Schr\"odinger equation tends to $-\frac{4\alpha_s^2 m_Q}{9}+c$. But according Fig.~\ref{fig:spectrum_QQ}, the dependence of the numerical results on $m_Q$ is not linear. The main reason is that $\alpha_s$ in the GIC model is not a constant. In addition, all potentials are smeared. In this scheme, the potential $\tilde{G}_{ij}$ in Eq.~(\ref{eq:smearG}) is not a strict Coulomb-type, but this result also provides us a good opportunity to study the spectroscopy behavior with large $m_Q$. We suggest precisely measuring the mass of $t\bar{t}$, which is crucial to test the effectiveness of the potential.

\section{The heavy-quark limit in top hadrons}\label{sec:HQS}
\begin{figure*}[htbp]
\begin{tabular*}{\textwidth}{@{\extracolsep{\fill}}ccc}
\includegraphics[width=0.310\textwidth]{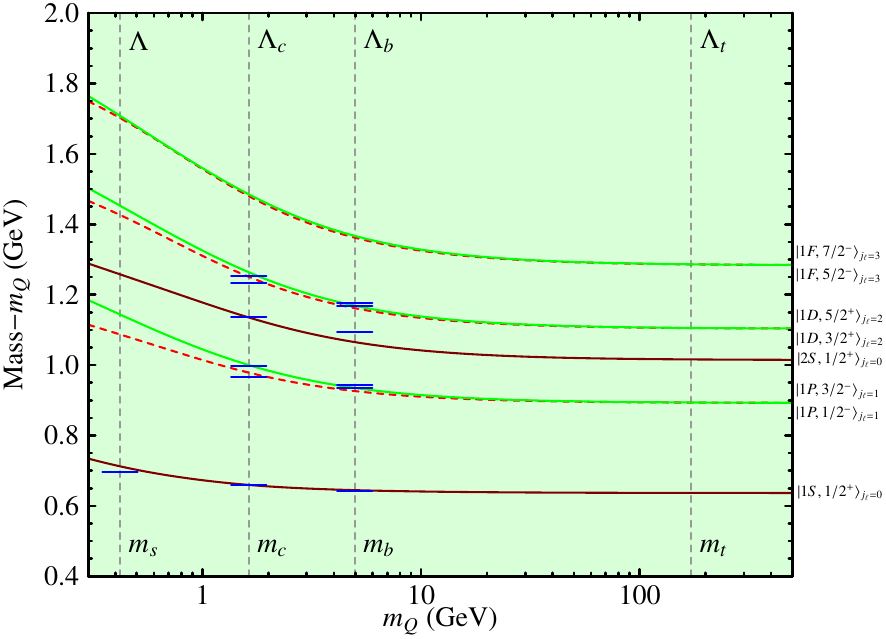}&
\includegraphics[width=0.310\textwidth]{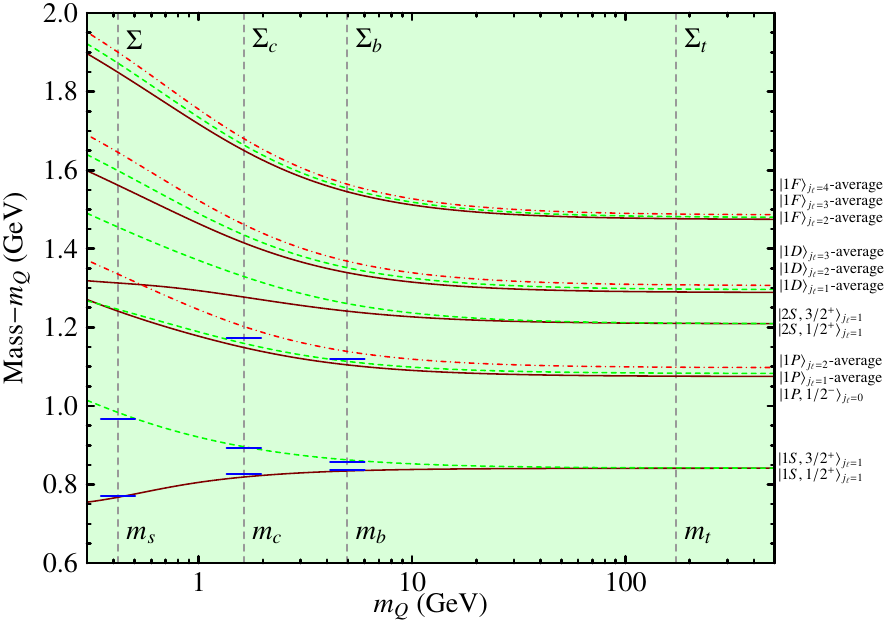}&
\includegraphics[width=0.310\textwidth]{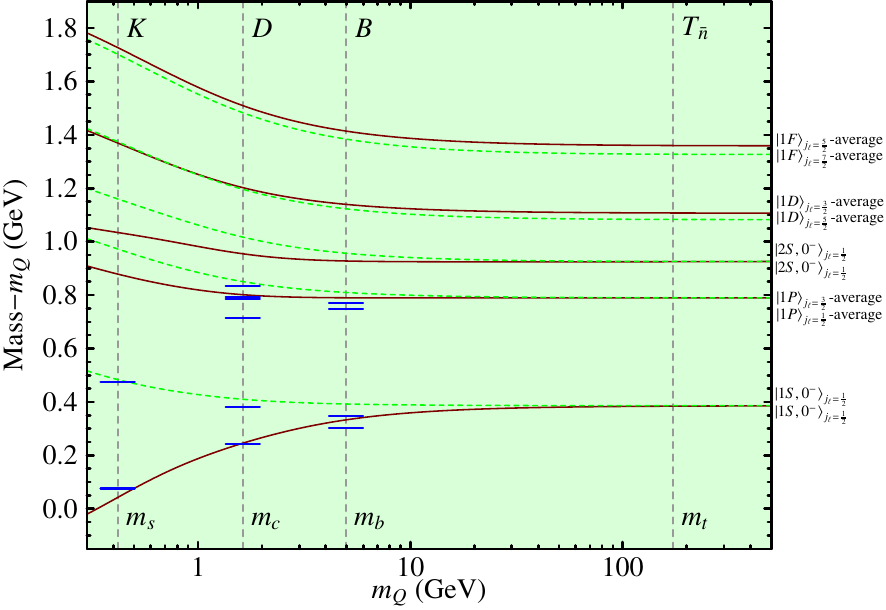}\\
(a)&(b)&(c)
\end{tabular*}
    \caption{The mass spectra of $\Lambda_Q$ (a), $\Sigma_Q$ (b), and heavy-light meson (c) with subtractions of heavy-quark mass $m_Q$. The blue short lines are experimental values. The $|NL\rangle_{j_\ell}$ implies we take the average mass of $|NL,j_\ell-\frac{1}{2}\rangle_{j_\ell}-|NL,j_\ell+\frac{1}{2}\rangle_{j_\ell}$}
    \label{fig:spectra}
\end{figure*}

As shown in Fig.~\ref{fig:spectra}, we calculate the mass spectra of $\Lambda_Q$, $\Sigma_Q$, and heavy-light mesons. In calculations, we change the heavy-quark mass $m_Q$ in the range of 0.3$\sim$500 GeV, which covers the charm, bottom, and topped consistent quark masses. It even includes the region of consistent strange quark mass. To reflect the relationships between the spectra and heavy-quark masses, the spectra were subtracted by heavy-quark mass $m_Q$.

With the latest data of the baryons, one can obtain the parameters of the potential model and then calculate the more excited baryons. According to Fig.~\ref{fig:spectra}, for $\Lambda$ and $\Sigma$, the $SU(3)$ may affect the spectroscopy properties, which make the excited modes more complex, but the the calculated ground masses of $\Lambda$ and $\Sigma^{(*)}$ highly match the measurements. In addition, for the singly heavy flavor baryons $\Lambda_c$, $\Lambda_b$, $\Sigma_c$, and $\Sigma_b$, the theoretical calculations are consistent with experimental results for both ground low-lying excited states. For the meson spectra, Godfrey and Isgur made a global fitting, i.e., from the light flavor meson $\pi$ to bottom quarkonia $b\bar{b}$. Here, we present the spectra of the $Q\bar{n}$-type meson and compare the experimental masses of the $K$, $D$, and $B$ mesons. These numerical results indicate that the GCI model can adapt to a wide variety of systems with different quark masses. However, for topped hadrons, since the topped hadrons were not observed in previous experiments, the applicability of the potential models was not verified. The new results from the CMS and ATLAS Collaborations may change the status~\cite{CMS:2025kzt,ATLAS:2026dbe}, where an enhancement structure near the $t\bar{t}$ threshold was observed, which is a possible pseudoscalar toponium. As discussed above, we calculate the masses of $^1S_0$ and $^3S_1$ toponia. The obtained masses are about 342.6 GeV, which is close to the threshold of $t\bar{t}$ and match the observations from the CMS and ATLAS Collaborations~\cite{CMS:2025kzt,ATLAS:2026dbe}. In this scheme, the potential model could not only be applied to the light flavor hadons, which include the $u$, $d$, and $s$ quarks, but they could also be applied to the hadrons that include the $c$, $b$, and the most massive quark $t$. 

In Fig.~\ref{fig:spectra}, we find that when we enlarge $m_Q$, the curves of the numerical results tend to flatten, whether ground or excited states. When $m_Q \to \infty$, the contribution of the kinetic energy 
term $\sqrt{m_Q^2+p_i^2}-m_Q$ is zero, the $\frac{1}{m_Q}$-dependent interactions also vanish, the semirelativistic correction terms have $\beta_{ij}\sim$1 and $\delta_{ij}\sim$1, and the parameters which depend on the $m_Q$ are also taken as the limits. Thus, in the heavy-quark limit, the mass spectra with the subtraction of $m_Q$ should be constants, which are reflected in the numerical results in Fig.~\ref{fig:spectra}. In previous works, we treated the charm and bottom quarks as heavy flavor quarks. Some spectrum and decay behaviors matched the heavy-quark symmetry. However, as shown in Fig.~\ref{fig:spectra}, $m_c$ is faraway the limit. The results of the bottom hadrons may be more close to the limit, but the bottom hadrons are not ideal systems when we study the heavy-quark symmetry. If we move our perspective to the top energy area, we find that the topped quark is a nearly perfect heavy flavor quark. For the topped hadrons, the spectra with subtraction of $m_Q$ of the ground and excited singly heavy flavor hadrons are very close to the limits, as presented in Fig.~\ref{fig:spectra}. First, the numerical results with $m_Q=m_t$ locate the nearly flat parts of the curves. Second, for the same $|NL\rangle_{j_\ell}$, the $J=\ell-\frac{1}{2}$ and $J=j_\ell+\frac{1}{2}$ have extremely similar masses, which could be read from the numerical results of the $S$-wave in Fig.~\ref{fig:spectra} and the orbital excited states of $\Lambda_Q$ in Fig.~\ref{fig:spectra}. The details of the orbital excited states of $\Sigma_t$ and $T_n$ can also be found in Tables~\ref{tab:tbaryon} and \ref{tab:tmeson}, respectively.

\section{Discussion and conclusions}\label{sec:summary}

The top quark occupies a unique position in the Standard Model due to its extremely large mass ($\sim 173$ GeV) and short lifetime ($\tau_t \sim 5\times10^{-25}$ s). Consequently, a fundamental question naturally arises whether it can form hadronic bound states before its decay, which makes the experimental observations of such systems seem impossible. The main reason for this concern is that the formation of a well-defined resonance or bound state  requires its formation timescale to be significantly shorter than its components' lifetime. In other words, for the lightest $t\bar{t}$ state, its estimated binding energy $E_B \sim \alpha_s^2 m_t $ is comparable in magnitude to $\Gamma_t \approx 1.4$ GeV. This implies that the formation timescale of a $t\bar{t}$ state is similar to the top quark lifetime, resulting in, at best, a broad, marginally resonant structure near the production threshold \cite{Strassler:1990nw}. 

However, the recent observation of an enhancement near the $t\bar{t}$ threshold by both the CMS and ATLAS experiments \cite{CMS:2025kzt,ATLAS:2026dbe} challenged this conventional understanding that no quasibound toponium states exist in nature, which immediately generated significant excitement in the scientific community, as it provided experimental support for predictions made by perturbative QCD and potential models \cite{Fadin:1987wz,Barger:1987xg,Kuhn:1987ty,Fadin:1990wx,Strassler:1990nw,Sumino:1997ve,Hoang:2000yr,Penin:2005eu,Hagiwara:2008df,Kiyo:2008bv,Sumino:2010bv,Beneke:2015kwa,Fuks:2021xje,Garzelli:2024uhe,Wang:2024hzd,Jiang:2024fyw,Akbar:2024brg,Fuks:2024yjj,Fu:2025yft}. In rapid sequence, a wide range of follow-up studies were carried out, including further investigations into spin correlations in $t\bar{t}$ production \cite{Hagiwara:2008df,Severi:2021cnj,Maltoni:2024tul,Aguilar-Saavedra:2024mnm,Nason:2025hix,Gombas:2025ibs}, advances in quantum theory \cite{Thompson:2025cgp,Lopez:2025kog,Shao:2025dzw}, potential explorations of new physics \cite{Butterworth:2025asm,LeYaouanc:2025mpk,Behring:2025ilo,CMS:2025dzq}, and discussions about the possibility of observing the toponium at future colliders \cite{Bai:2025buy,Xiong:2025iwg,Maltoni:2024csn,Maltoni:2024}.

We also point out that the topped spectroscopy is crucial to test the applicability of the potential model. Previous experience has shown that the potential models could be well employed in the light flavor, charm, and bottom hadrons, including both mesons and baryons. Currently, it remains unknown whether the potential model can be extended infinitely as quark masses increase. The topped hadron is a good platform to decode this problem. The observation of the $t\bar{t}$ enhancement provides us a good chance to investigate it. The calculations imply that the theoretical masses of the quarkonia match the experimental results pretty well with the quark masses below that of the topped quark. If the calculated masses of the toponium are also consistent with the measurements, one could promote the potential model to a topped scenario. In this way, the precise measurement of the $t\bar{t}$ enhancement mass plays a crucial role in testing the potential model. We also notice that there exist a obvious mass gaps between topped and other quarks. The large mass makes the topped quark a nearly ideal heavy-quark, whose phenomena appear in the singly heavy flavor systems when we enlarge the mass of the heavy flavor quark. Different from the very large binding energy of the $t\bar{t}$ systems, the mass of a singly heavy flavor hadron tends to a constant in the same quantum number.

\begin{acknowledgments}
This work is supported by the National Natural Science Foundation of China under Grant Nos. 12335001, 12305087, 12247101, and 12405098,  the ‘111 Center’ under Grant No. B20063, the Natural Science Foundation of Gansu Province (No. 26RCKA012, No. 22JR5RA389, No. 25JRRA799), the Talent Scientific Fund of Lanzhou University, the fundamental Research Funds for the Central Universities (No. lzujbky-2023-stlt01), the project for top-notch innovative talents of Gansu province, and Lanzhou City High-Level Talent Funding, and the Start-up Funds of Nanjing Normal University under Grant No.~184080H201B20.
\end{acknowledgments}

\section*{DATA AVAILABILITY}

The data supporting the findings of this study is available on Zenodo repository~\cite{Luo:2026top}.

\bibliographystyle{UserDefined}
\bibliography{references}

\end{document}